\documentclass[preprint, superscriptaddress, showpacs,preprintnumbers,amsmath,amssymb,pra]{revtex4}
\usepackage{graphicx}

\begin{document}

\thispagestyle{empty}

\title{Observability of thermal effects in the Casimir interaction from
 graphene-coated substrates}

\author{
G.~L.~Klimchitskaya}
\affiliation{Central Astronomical Observatory at Pulkovo of the Russian Academy of Sciences, St.Petersburg,
196140, Russia}
\affiliation{Institute of Physics, Nanotechnology and
Telecommunications, St.Petersburg State
Polytechnical University, St.Petersburg, 195251, Russia}

\author{
V.~M.~Mostepanenko}
\affiliation{Central Astronomical Observatory at Pulkovo of the Russian Academy of Sciences, St.Petersburg,
196140, Russia}
\affiliation{Institute of Physics, Nanotechnology and
Telecommunications, St.Petersburg State
Polytechnical University, St.Petersburg, 195251, Russia}

\begin{abstract}
Using the recently proposed theory, we calculate thermal
effect in the Casimir interaction from graphene-coated
metallic and dielectric substrates. The cases when only
one or both of the two parallel plates are coated with
graphene are considered. It is shown that the graphene
coating does not influence the Casimir interaction between
metals, but produces large impact for dielectrics. This
impact increases with decreasing static dielectric
permittivity of the plate material. The thermal correction
to the gradient of the Casimir force between an Au sphere
and graphene coated fused silica plate is calculated.
It is shown to be significanlty greater than the total
experimental error in the recently performed experiment,
which is demonstrated to be only one step away from
observation of the thermal effect from a graphene-coated
substrate at short separation distances. To achieve this
goal, one should increase the thickness of the fused silica
film from 300\,nm to $2\,\mu$m.
\end{abstract}

\maketitle

\section{Introduction}

Thermal effects in the Casimir interaction have been the subject
of much attention (see Ref.~\cite{1} for a review).
However, until the present day they are not observed at short
separations of the order of 100\,nm, i.e., in the region
characterized by the highest experimental precision.
In this respect such unusual material
as graphene, which is a 2D sheet of
carbon atoms, is of special interest.
As was shown in Ref.~\cite{2}, for graphene the thermal effects
become crucial at much shorter separations than for ordinary
materials. This result was obtained by using the longitudinal
density-density correlation function and later reconfirmed
within quantum electrodynamical formalism of the
polarization tensor in (2+1)-dimensional space-time
\cite{3,4}. In doing so, graphene was described in the framework
of the Dirac model \cite{5} which is applicable at low energies
below several eV,
as appropriate for fluctuation-induced van der Waals and Casimir
forces. Later on, a lot of calculations of graphene-graphene,
graphene-plate, and atom-graphene Casimir and Casimir-Polder
interactions has been performed using different
methods \cite{6,7,8,9,10,11,12,13,14,15,16}.

Measurements of the Casimir force between two freestanding
graphene sheets
or between a graphene sheet and a material plate present a
real challenge due to deformations of graphene surfaces.
Because of this, in the pioneering experiment \cite{17} it was
chosen to measure the gradient of the Casimir force between an
Au-coated sphere and a graphene sheet deposited on a SiO${}_2$
film covering a Si plate. Measurements were performed using a
dynamic atomic force microscope (AFM) operated in the frequency
shift mode \cite{17}. However, the comparison of the
measurement data with theory was made difficult because the
available reflection coefficients for graphene deposited on a
substrate \cite{18,19,20} were expressed in terms of the
dielectric permittivity of substrate material and the
density-density correlation functions of graphene.
The main difficulty encounted in computations was that the
expression for the transverse density-density correlation
function, as well as the explicit temperature-dependence of
both longitudinal and transverse functions, remained unknown.
In these conditions, Ref.~\cite{17} used the additive theory
which overestimates the force gradient, and, as a consequence,
only a qualitative agreement with the experimental data was
achieved.

The state of affairs
 has been changed only recently due to two important
results obtained in the literature. In Ref.~\cite{21}, an
explicit connection between the density-density correlation
functions and the components of the polarization tensor was
established. In the process, the analytic expression for the
transverse correlation function and the temperature dependence
of both longitudinal and transverse functions have been found.
Furthermore, in Ref.~\cite{22} the reflection coefficients for
a graphene sheet deposited on a substrate were derived
independently in terms of the
dielectric permittivity of a substrate material and the
polarization tensor of graphene. The results of
Refs.~\cite{21,22} were found to be in agreement.
On this basis the experimental data of Ref.~\cite{17} were
compared with the proposed theory and a very good
quantitative agreement was demonstrated within the limits
of the experimental errors. This raises a question as to
whether thermal effects in Casimir interaction between
graphene-coated substrates are observable using the existing
laboratory setup.

In this paper, we calculate the thermal Casimir pressure
between metallic and dielectric substrates coated with graphene.
The cases when only one or both two material plates are coated
with a graphene sheet are considered. We perform computations for
gold, silicon, sapphire, mica and fused silica substrates.
It is shown that the role of graphene is the most pronounced for
a substrate material having the smallest static dielectric
permittivity (fused silica in our case). For metallic substrates
the deposition of a graphene sheet does not lead to a detectable
change in the Casimir  pressure. Then we calculate the thermal
correction to the gradient of the Casimir force between an Au
sphere and a SiO${}_2$ plate coated with a graphene sheet at
the experimental separations of a few hundred nanometers.
It is shown that over a sufficiently wide region of separations
the thermal correction is up to a factor of 5 larger than the
total experimental error. Finally, we demonstrate that the
experiment of Ref.~\cite{17}, which was the pioneer measurement
of the Casimir interaction with a graphene-coated substrate,
was only one step away from measuring the respective thermal
effect as well. According to our results, the latter goal can
be achieved if to increase the thickness of SiO${}_2$ film
in Ref.~\cite{17} from 300\,nm to $2\,\mu$m.

The paper is organized as follows. In Sec.~II we briefly present
the Lifshitz formula for graphene-coated substrates where the
reflection coefficients are expressed in terms of the dielectric
permittivity and the polarization tensor. The influence of graphene
films on the thermal Casimir pressure between material plates is
analyzed in Sec.~III. In Sec.~IV the thermal effect
in the gradient of the Casimir force between an Au-coated sphere
and a SiO${}_2$ plate is calculated
and the possibility to observe it is demonstrated. Section~V
contains our conclusions and discussion.

\section{The Lifshitz formula for graphene-coated substrates}

In this section we consider the thermal Casimir pressure
$P(a,T)$ between two thick plates (semispaces) separated by a
distance $a$ at least one of which is coated with a graphene
sheet. The Lifshitz formula in terms of the reflection
coefficients takes the standard form \cite{23,24}
\begin{eqnarray}
&&
P(a,T)=-\frac{k_BT}{8\pi a^3}\sum_{l=0}^{\infty}
{\vphantom{\sum}}^{\prime}\int_{\zeta_l}^{\infty}\!\!y^2dy
\left[\frac{R_{\rm TM}^{(1)}(i\zeta_l,y)
R_{\rm TM}^{(2)}(i\zeta_l,y)}{e^{y}-
R_{\rm TM}^{(1)}(i\zeta_l,y)R_{\rm TM}^{(2)}(i\zeta_l,y)}
\right.
\nonumber \\
&&~~~~~~
\left.
+\frac{R_{\rm TE}^{(1)}(i\zeta_l,y)
R_{\rm TE}^{(2)}(i\zeta_l,y)}{e^{y}-
R_{\rm TE}^{(1)}(i\zeta_l,y)R_{\rm TE}^{(2)}(i\zeta_l,y)}
\right].
\label{eq1}
\end{eqnarray}
\noindent
Here, $k_B$ is the Boltzmann constant, $T$ is the temperature,
and the dimensionless Matsubara frequencies $\zeta_l$ are
expressed via the dimensional ones by
$\zeta_l=2a\xi_l/c$, where $\xi_l=2\pi k_BTl/\hbar$ with
$l=0,\,1,\,2,\,\ldots$ (the prime on the summation sign indicates
that the term with $l=0$ is divided by two).
Note that the dimensionless variable $y$ is connected by the
equation
\begin{equation}
y=2aq_l\equiv2a\left(k_{\bot}^2+\frac{\xi_l^2}{c^2}
\right)^{1/2}
\label{eq2}
\end{equation}
\noindent
with the projection of the wave vector on the plane of
plates $k_{\bot}$.

The reflection coefficients on the graphene-coated substrates
were found by different methods in Refs.~\cite{18,19,20}
combined with Ref.~\cite{21},
on the one hand,  and in Ref.~\cite{22}, on the other hand.
They are given by
\begin{eqnarray}
&&
R_{\rm TM}^{(n)}(i\zeta_l,y)=\frac{\varepsilon_l^{(n)}y+
k_l^{(n)}\left(\frac{y}{y^2-\zeta_l^2}\tilde{\Pi}_{00}-1
\right)}{\varepsilon_l^{(n)}y+k_l^{(n)}
\left(\frac{y}{y^2-\zeta_l^2}\tilde{\Pi}_{00}+1
\right)},
\nonumber \\[-1mm]
&&\phantom{aaa}
\label{eq3} \\[-2mm]
&&
R_{\rm TE}^{(n)}(i\zeta_l,y)=\frac{y-k_l^{(n)}-
\left(\tilde{\Pi}_{\rm tr}-\frac{y^2}{y^2-\zeta_l^2}
\tilde{\Pi}_{00}\right)}{y+k_l^{(n)}+
\left(\tilde{\Pi}_{\rm tr}-\frac{y^2}{y^2-\zeta_l^2}
\tilde{\Pi}_{00}\right)},
\nonumber
\end{eqnarray}
\noindent
where the dielectric permittivities for the materials
of the first and second semispaces ($n=1,\,2$) are
\begin{equation}
\varepsilon_l^{(n)}\equiv
\varepsilon^{(n)}(i\xi_l)=
\varepsilon^{(n)}(ic\zeta_l/2a),
\label{eq3a}
\end{equation}
\noindent
$\tilde{\Pi}_{00}$ is the 00-component,
$\tilde{\Pi}_{\rm tr}$ is the sum of spatial components
$\tilde{\Pi}_{1}^{\,1}$ and $\tilde{\Pi}_{2}^{\,2}$
of the dimensionless polarization tensor in
(2+1)-dimensional space time \cite{3,4}, and the following
notation is introduced
\begin{equation}
k_l^{(n)}\equiv
\sqrt{y^2+(\varepsilon_l^{(n)}-1)\zeta_l^2}.
\label{eq3b}
\end{equation}
\noindent
The dimensionless components are connected with the
dimensional ones by the equations
\begin{equation}
\tilde{\Pi}_{00}\equiv\tilde{\Pi}_{00}(i\zeta_l,y)=
\frac{2a}{\hbar}\Pi_{00},
\qquad
\tilde{\Pi}_{\rm tr}\equiv
\tilde{\Pi}_{\rm tr}(i\zeta_l,y)=
\frac{2a}{\hbar}\Pi_{\rm tr}.
\label{eq4}
\end{equation}

Now we present analytic expressions for the components of the
polarization tensor entering Eq.~(\ref{eq3}).
This tensor depends on $T$ both explicitly as on a parameter and
implicitly through the Matsubara frequencies.
It was shown \cite{4,10,12,13} that an explicit dependence
on $T$ substantially affects the computational results for the
Casimir free energy and pressure only through the zero-frequency
term of the Lifshitz formula $l=0$. In so doing, all terms of
Eq.~(\ref{eq1}) with $l\geq 1$ without the loss of accuracy can
be calculated using the polarization tensor defined at $T=0$
and depending on $T$ only implicitly.

The respective exact expressions at $\zeta_0=0$ for a graphene
with a nonzero mass gap parameter $\Delta$ are \cite{4,10}
\begin{eqnarray}
&&
\tilde{\Pi}_{00}(0,y)=\frac{8\alpha}{\tilde{v}_F^2}\left[
\frac{\tau}{\pi}\int_{0}^{1}\!\!dx\ln\left(2
\cosh\frac{\pi\theta}{\tau}\right)
-\tilde{\Delta}^{\!2}\int_{0}^{1}\frac{dx}{\theta}
\tanh\frac{\pi\theta}{\tau}\right],
\label{eq5} \\
&&
\tilde{\Pi}_{\rm tr}(0,y)-\tilde{\Pi}_{00}(0,y)=
8\alpha\tilde{v}_F^2y^2\!\int_{0}^{1}\!\!\!dx
\frac{x(1-x)}{\theta}
\tanh\frac{\pi\theta}{\tau}.
\nonumber
\end{eqnarray}
\noindent
Here, $\alpha=e^2/(\hbar c)$ is the fine-structure constant,
$\tilde{v}_F=v_F/c$, where $v_F\approx 9\times 10^5\,$m/s
is the Fermi velocity in graphene \cite{25,26},
$\tau\equiv 4\pi ak_BT/(\hbar c)$ is the temperature parameter,
$\tilde{\Delta}\equiv 2a\Delta/(\hbar c)$, and the function
$\theta$ is defined as
\begin{equation}
\theta\equiv\theta(x,y)=[\tilde{\Delta}^{\!2}+
x(1-x)\tilde{v}_F^2y^2]^{1/2}.
\label{eq6}
\end{equation}
\noindent
It is seen that the right-hand sides
in Eq.~(\ref{eq5})  depend on $T$ explicitly
through the dimensionless parameter $\tau$.

At all Matsubara frequencies $\zeta_l$ with $l\geq 1$ one can
use the following expressions for the components of the
polarization tensor at $T=0$, where the continuous $\zeta$ are
replaced with the discrete $\zeta_l$ \cite{3,4,10}:
\begin{eqnarray}
&&
\tilde{\Pi}_{00}(i\zeta_l,y)=\alpha
\frac{y^2-\zeta_l^2}{f^2(\zeta_l,y)}
\,\Phi(\zeta_l,y),
\label{eq7} \\
&&
\tilde{\Pi}_{\rm tr}(i\zeta_l,y)-
\frac{y^2}{y^2-\zeta_l^2}\tilde{\Pi}_{00}(i\zeta_l,y)=
\alpha\Phi(\zeta_l,y),
\nonumber
\end{eqnarray}
\noindent
where the functions $\Phi$ and $f$ are given by
\begin{eqnarray}
&&
f(\zeta_l,y)=[\tilde{v}_F^2y^2+
(1-\tilde{v}_F^2)\zeta_l^2]^{1/2},
\label{eq8} \\
&&
\Phi(\zeta_l,y)=4\tilde{\Delta}+2f(\zeta_l,y)\left[1-
\frac{4\tilde{\Delta}^{\!2}}{f^2(\zeta_l,y)}\right]
\arctan\frac{f(\zeta_l,y)}{2\tilde{\Delta}}.
\nonumber
\end{eqnarray}

The reflection coefficients between thick plates (semispaces)
made of ordinary materials with no graphene coatings are
obtained from Eq.~(\ref{eq3}) by putting
$\tilde{\Pi}_{00}=\tilde{\Pi}_{\rm tr}=0$.
If the graphene coating is missing on only one plate,
$n=2$, for instance, the polarization tensor should be put
equal to zero only in these specific reflection coefficients
$R_{\rm TM}^{(2)}$ and $R_{\rm TE}^{(2)}$.

\section{Influence of graphene coatings on the thermal
Casimir pressure}

In this section we calculate the influence of graphene coatings
on the Casimir pressure at $T=300\,$K for plates (semispaces)
made of various materials, both metallic and dielectric,
using the Lifshitz formula (\ref{eq1}) and the reflection
coefficients (\ref{eq3}).  For the sake of simplicity,
here we consider the case of pristine (gapless) graphene
because at $T=300\,$K the allowable nonzero mass gap
parameters ($\Delta<0.1\,$eV \cite{3}) lead to only minor
deviations in the magnitudes of the thermal Casimir force,
as compared to the case $\Delta=0$ \cite{10}.
In Sec.~IV the influence of nonzero mass gap parameter
is specified in more detail.

We examine the case when both plates are made of a common
material, so that
$\varepsilon_l^{(1)}=\varepsilon_l^{(2)}$, and either one of
them or both two are coated with a graphene sheet.
As the first example, we consider Au plates.
The dielectric permittivity of Au at the imaginary Matsubara
frequencies is obtained by means of
the Kramers-Kronig relation
from the tabulated optical data
\cite{27} extrapolated down to zero frequency \cite{1,24}.
Computations show that the values of the ratios
$P_g/P$ and $P_{gg}/P$, where $P_g$ and $P_{gg}$ are the
Casimir pressures between two Au plates one of which or both
two are coated with graphene, respectively, and $P$ is the
pressure between uncoated plates, do not depend on the type
of extrapolation of the optical data by means of the Drude
or the plasma model. The values of the plasma frequency
$\omega_p=9.0\,$eV and the relaxation parameter
$\gamma=0.035\,$eV have been used in extrapolations.

The computational results for the ratios
$P_g/P$ and $P_{gg}/P$ are shown by the lines numbered 1
in Fig.~1(a) and Fig.~1(b), respectively, as functions of
separation. As can be seen in these figures, the Casimir
pressure between Au plates is scarcely affected by graphene
coatings over the wide range of separations from 100\,nm to
$6\,\mu$m. As an illustration, at $a=100\,$nm we have
$P_g/P=1.0013$ and $P_{gg}/P=1.0025$, and both ratios
quickly decrease to unity with increasing separation. The same holds
for any metal or even for a doped semiconductor in a metallic
state (i.e., for the concentration of charge carriers above
the critical value). Thus, for B-doped Si (the concentration
of charge carriers
$n\approx 1.6\times 10^{19}\,\mbox{cm}^{-3}$) at $a=100\,$nm
one obtains
$P_g/P=1.0041$ and $P_{gg}/P=1.0082$, i.e., the
maximum influence of graphene coatings is less than 1\%.

To obtain the larger influence of graphene coatings on the
Casimir pressure, we consider substrates made of different
dielectric materials, such as high-resistivity Si,
sapphire (Al${}_2$O${}_3$), mica, and fused silica
(SiO${}_2$) possessing the static dielectric permittivities
$\varepsilon(0)$ equal to 11.7, 10.1, 5.4, and 3.8,
respectively. The dielectric permittivity of Si along the
imaginary frequency axis was obtained from the tabulated
optical data \cite{28} by means of the Kramers-Kronig
relation. As to sapphire, mica, and fused silica, the
available sufficiently precise analytic representations
for their frequency-dependent dielectric permittivities
have been used \cite{29}.

The computational results for the ratios
$P_g/P$ and $P_{gg}/P$ as functions of
separation are shown in Fig.~1(a) and Fig.~1(b), respectively,
by the lines 2 (high-resistivity Si), 3 (sapphire), 4 (mica),
and 5 (fused silica). As is seen in Fig.~1, for dielectric
plates the presence of graphene coatings significantly
influences the Casimir pressure. This influence is larger
when both dielectric plates are coated with graphene
comparing to the case when only one plate is graphene-coated.
It is significant that influence of graphene coatings on the
Casimir pressure increases with decreasing static dielectric
permittivity of the substrate material. This is the case for
both one and two plates coated with graphene. {}From Fig.~1
we conclude that the largest impact of graphene coatings on
the Casimir pressure takes place for the fused silica substrate.
Thus, when one of the two fused silica plates is coated with
graphene we have $P_g/P=1.17$, 1.25, 1.43, and 1.78 at
separation distances $a=200\,$nm, 400\,nm, $1\,\mu$m, and
$6\,\mu$m, respectively. For two graphene-coated fused silica
plates $P_{gg}/P=1.47$, 1.72, 2.28, and 3.34 at the same
respective separations.
Thus, for dielectric substrates the influence of graphene coatings
can be sufficiently large even at short separation distances below
$1\,\mu$m, i.e., in the measurement region of precise
experiments on measuring the Casimir interaction (see
Ref.~\cite{1} and more recent experiments \cite{17,30,31,32,33}).

To understand the absolute value of the calculated effects, in
Fig.~2 we plot as functions of separation the magnitudes of the
Casimir pressure between two fused silica plates (the dashed
line), between one graphene-coated and one uncoated (the lower
solid line) and between two graphene-coated fused silica plates
(the upper solid line). For better visualization, the separation
region from 100\,nm to $1\,\mu$m is shown on an inset.
As can be seen in Fig.~2, the presence of graphene layers
essentially changes the magnitudes of the Casimir pressure
both at short and long separation distances.
Taking into account that the thermal effect in graphene is a
special case, this can be used for its observation in
measuring the Casimir interaction between graphene-coated
substrates.
We consider this possibility in the next section.

\section{Possibility to observe the thermal effect in
Casimir interaction with graphene-coated substrates}

Precise measurements of the Casimir interaction mentioned
above were performed in the configuration of a sphere of
radius $R$ above a plane plate. The separation distance between
the sphere and the plate was always much smaller than $R$, i.e.,
the inequality $a\ll R$ was satisfied with a wide safety
margin. Thus, the comparison of the measurement data with theory
can be performed using the proximity force approximation (PFA)
stating that the Casimir force between the sphere and the plate
is given by \cite{1,24}
\begin{equation}
F_{\!\!sp}(a,T)=2\pi R {\cal F}(a,T),
\label{eq9}
\end{equation}
\noindent
where ${\cal F}(a,T)$ is the free energy of the Casimir
interaction between two parallel plates per unit area.
Calculating the negative derivative of Eq.~(\ref{eq9}) with
respect to $a$, one obtains a connection between the
measured force gradient in sphere-plate geometry and the
effective Casimir pressure between two parallel plates
\begin{equation}
\frac{F_{\!\!sp}^{\prime}(a,T)}{R}=-2\pi P(a,T).
\label{eq10}
\end{equation}
\noindent
Note that using the exact theory of the Casimir interaction,
applicable to boundary surfaces of arbitrary shape, the relative
error of Eq.~(\ref{eq10}) was shown to be smaller than
$(0.3\div 0.4)a/R$ \cite{34,35,36,37}.

Now we consider the proposed experimental configuration of
an Au-coated sphere and a graphene-coated plate made of
SiO${}_2$, which is a substrate leading to the largest
influence of graphene coating (see Sec.~III).
Although according to the results of Sec.~III the coating of
both test bodies by graphene influences the Casimir pressure
even stronger, this option is not feasible experimentally
because the second body is of spherical shape.

We have computed the quantity $F_{\!\!sp}^{\prime}/R$ for an Au
sphere interacting with a graphene-coated SiO${}_2$ plate using
Eqs.~(\ref{eq1}) and (\ref{eq10}). The reflection coefficients
$R_{\rm TM,TE}^{(1)}$ in Eq.~(\ref{eq3}) are calculated with the
dielectric permittivity $\varepsilon_l^{(1)}$ of SiO${}_2$
and the polarization tensor of graphene. The reflection
coefficients $R_{\rm TM,TE}^{(2)}$  are calculated with the
dielectric permittivity $\varepsilon_l^{(2)}$ of Au and
$\tilde{\Pi}_{00}=\tilde{\Pi}_{\rm tr}=0$.
The type of extrapolation of the optical data for Au to
zero frequency leads to only a minor influence
on the computational results
\cite{10}. The computations were performed at $T=300\,$K and
at $T=0\,$K. In the latter case the summation in Eq.~(\ref{eq1})
was replaced with an integration over continuous frequency
\cite{24}. The computational results as functions of separation
are shown in Fig.~3(a) by the solid and dashed lines for
$T=300\,$K and at $T=0\,$K, respectively.

The thermal correction to the force gradient normalized by $R$,
\begin{equation}
\frac{1}{R}{\Delta_TF_{\!\!sp}^{\prime}(a)}\equiv
\frac{1}{R}\left[F_{\!\!sp}^{\prime}(a,T=300\,\mbox{K})-
F_{\!\!sp}^{\prime}(a,T=0\,\mbox{K})\right],
\label{eq11}
\end{equation}
\noindent
is shown by the line 1 in Fig.~3{b} as a function of separation.
The solid line 2 indicates the value of the total experimental
error in measurements of the normalized force gradient in the
experiment of Ref.~\cite{17} equal to 0.012\,Pa.
As is seen in Fig.~3(b), the thermal correction markedly exceeds
the total experimental error at separations below 350\,nm.
In Fig.~3(c), the relative thermal correction
\begin{equation}
\delta_TF_{\!\!sp}^{\prime}(a)=
\frac{\Delta_TF_{\!\!sp}^{\prime}(a)}{F_{\!\!sp}^{\prime}(a,
T=300\,\mbox{K})}
\label{eq12}
\end{equation}
\noindent
is shown by the line 1 as a function of separation.
In the same figure the solid line 2 indicates the relative
error in measurements of the force gradient.
It is again seen that the thermal effect due to the graphene
coating of the SiO${}_2$ plate is observable using the
parameters of already existing laboratory setup \cite{17}.

Now we discuss whether the thermal correction due to graphene
sheet deposited on a substrate was observed in the experiment
of Ref.~\cite{17}. In this experiment, the gradient of the
Casimir force was measured between an Au-coated sphere of
$R=54.1\,\mu$m radius and a graphene sheet deposited on a
SiO${}_2$ film of thickness $D=300\,$nm covering a
B-doped Si plate of thickness $500\,\mu$m.
The latter can be considered as a semispace.
As was mentioned in Sec.~I, in Ref.~\cite{22} the measurement
data of Ref.~\cite{17} were compared with the theory describing
graphene-coated substrates and a very good agreement was found.
This comparison was performed at the laboratory temperature
$T=300\,$K. The computations at $T=0\,$K were not performed,
and the possibility to observe the thermal effect was not
discussed.

The gradient of the Casimir force at $T=0$ can be computed
using Eqs.~(\ref{eq10}) and (\ref{eq1}). In the latter the
discrete summation should be replaced with an integration over
continuous frequency. The reflection coefficients
$R_{\rm TM,TE}^{(2)}$ on an Au semispace are expressed as
discussed above. The reflection coefficients
$R_{\rm TM,TE}^{(g,f,s)}$ on a graphene sheet deposited on a
SiO${}_2$ film covering Si semispace, which should be used
instead of $R_{\rm TM,TE}^{(1)}$ in Eq.~(\ref{eq1}), are
expressed by using the standard formulas of the Lifshitz theory
for planar layered structures \cite{24,38,39}
\begin{equation}
R_{\rm TM,TE}^{(g,f,s)}(i\zeta,y)=
\frac{R_{\rm TM,TE}^{(1)}(i\zeta,y)+
r_{\rm TM,TE}^{(f,s)}(i\zeta,y)e^{-2Dk_f^{(1)}}}{1+
R_{\rm TM,TE}^{(1)}(i\zeta,y)
r_{\rm TM,TE}^{(f,s)}(i\zeta,y)e^{-2Dk_f^{(1)}}}.
\label{eq13}
\end{equation}
\noindent
Here, the coefficients $R_{\rm TM,TE}^{(1)}$ describe the
reflection on a graphene sheet deposited on a SiO${}_2$
semispace. They are given in Eq.~(\ref{eq3}) where
$\varepsilon^{(1)}\equiv\varepsilon_f^{(1)}$ is the dielectric
permittivity of SiO${}_2$ along the imaginary frequency axis.
The coefficients $r_{\rm TM,TE}^{(f,s)}$ are well known Fresnel
coeffisients. They describe the reflection on the boundary
between two semispaces made of SiO${}_2$ and Si
\begin{equation}
r_{\rm TM}^{(f,s)}(i\zeta,y)=
\frac{\varepsilon_s^{(1)}k_f^{(1)}-
\varepsilon_f^{(1)}k_s^{(1)}}{\varepsilon_s^{(1)}k_f^{(1)}+
\varepsilon_f^{(1)}k_s^{(1)}},
\qquad
r_{\rm TE}^{(f,s)}(i\zeta,y)=
\frac{k_f^{(1)}-k_s^{(1)}}{k_f^{(1)}+k_s^{(1)}}.
\label{eq14}
\end{equation}
\noindent
The quantity $\varepsilon_s^{(1)}$ is the dielectric permittivity
of Si calculated along the imaginary frequency axis, and
$k_s^{(1)}$, $k_f^{(1)}$ are defined in Eq.~(\ref{eq3b}) and
calculated with respective dielectric permittivities
$\varepsilon_s^{(1)}$ and $\varepsilon_f^{(1)}$.

In Fig.~4 the measurement data of Ref.~\cite{17} are indicated as
crosses. The arms of the crosses show the errors in the measured
separation distances and force gradients determined at the 67\%
confidence level.
The results of theoretical computations made in Ref.~\cite{22}
at $T=300\,$K using the proposed theory describing
graphene-coated substrates are shown as the light-gray band.
The width of the band is determined by the uncertainty in the
plasma frequency of metallic Si used in Ref.~\cite{17},
differences between the predictions of the Drude and plasma
model extrapolations of the optical data of Au and Si to zero
frequency, and by the uncertainty of the mass gap parameter of
graphene within the interval from 0 to 0.1\,eV.
In the same figure, our computational results at $T=0\,$K are
presented by the dark-gray band. The larger thickness of this
band is explained by the fact that the influence
of nonzero mass gap parameter $\Delta$ on the computational
results is stronger at $T=0\,$K.
Note that the results of computations at both  $T=300\,$K and
$T=0\,$K were corrected for the presence of surface roughness
measured by means of AFM. This correction, however, was found
to be below 0.1\% of the calculated force gradients.
As can be seen in Fig.~4, the computational results at
$T=300\,$K are in a very good agreement with the measurement
data, whereas the computations using the same theory, but
performed at  $T=0\,$K, deviate from the data by touching
them only slightly. This means that the experiment of
Ref.~\cite{17} was only one step away from measuring the
thermal effect between an Au sphere and a graphene coated
substrate. According to our resulys, the single point in
this experiment which needs to be revised is the increased
up to $2\,\mu$m thickness of the SiO${}_2$ film.
We have checked by means of numerical computations that in
this case the SiO${}_2$ film below graphene can be already
considered as a semispace, and the underlying Si plate has
no detrimental effect by decreasing the thermal correction
to the measured force gradient.

\section{Conclusions and discussion}

In this paper we have investigated the thermal effect in the
Casimir interaction between graphene-coated substrates using the
recently proposed theory. The thermal Casimir pressure as a
function of separatiom was calculated when only one or both of
the two parallel plates are coated with a graphene sheet.
As the plate materials, we have considered Au or other metals
and also different dielectrics (dielectric silicon, sapphire,
mica and fused silica).
It was shown that the graphene coating of metallic substrate
does not influence the thermal Casimir pressure.
For dielectric materials the influence of graphene coating
is shown to increase with decreasing  static dielectric
permittivity of substrate material. Thus, among the materials
mentioned above, we have the largest impact of graphene
coating on the Casimir pressure in the case of fused silica.

Furthermore, we have calculated both the absolute and
relative thermal corrections to the gradient of the Casimir
force between an Au sphere and graphene-coated fused silica
plate, which is the configuration of recent experiment
\cite{17}. According to our result, at separations below
350\,nm the magnitude of the thermal correction up to a
factor of 5 exceeds the total experimental error in the
measured force gradient. This means that, when employing the
appropriate substrate material, it is possible to observe
the thermal effect from deposition of graphene sheet by
using the already existing experimental setup.

Finally, we have calculated the zero-temperature gradient
of the  Casimir force in the experiment \cite{17}, i.e.,
between an Au sphere and a graphene sheet deposited on a
SiO${}_2$ film covering a Si plate. It was shown that the
computational results at $T=0\,$K deviate from the
measurement data by touching them only slightly.
The same data are in a very good agreement with the
computational results at $T=300\,$K \cite{22}.
It is concluded that the experiment \cite{17} was only
one step away from measuring the thermal effect from
graphene deposited on a SiO${}_2$ substrate.
To achieve this goal, it would be necessary to increase
the thickness of a SiO${}_2$ film from 300\,nm to
$2\,\mu$m.

To conclude, we have shown that thermal effects in the
Casimir interaction from graphene-coated substrates are
observable at short separations below $1\,\mu$m, where
the highest experimental precision is achieved.


\begin{figure}[b]
\vspace*{-2cm}
\centerline{\hspace*{1cm}
\includegraphics{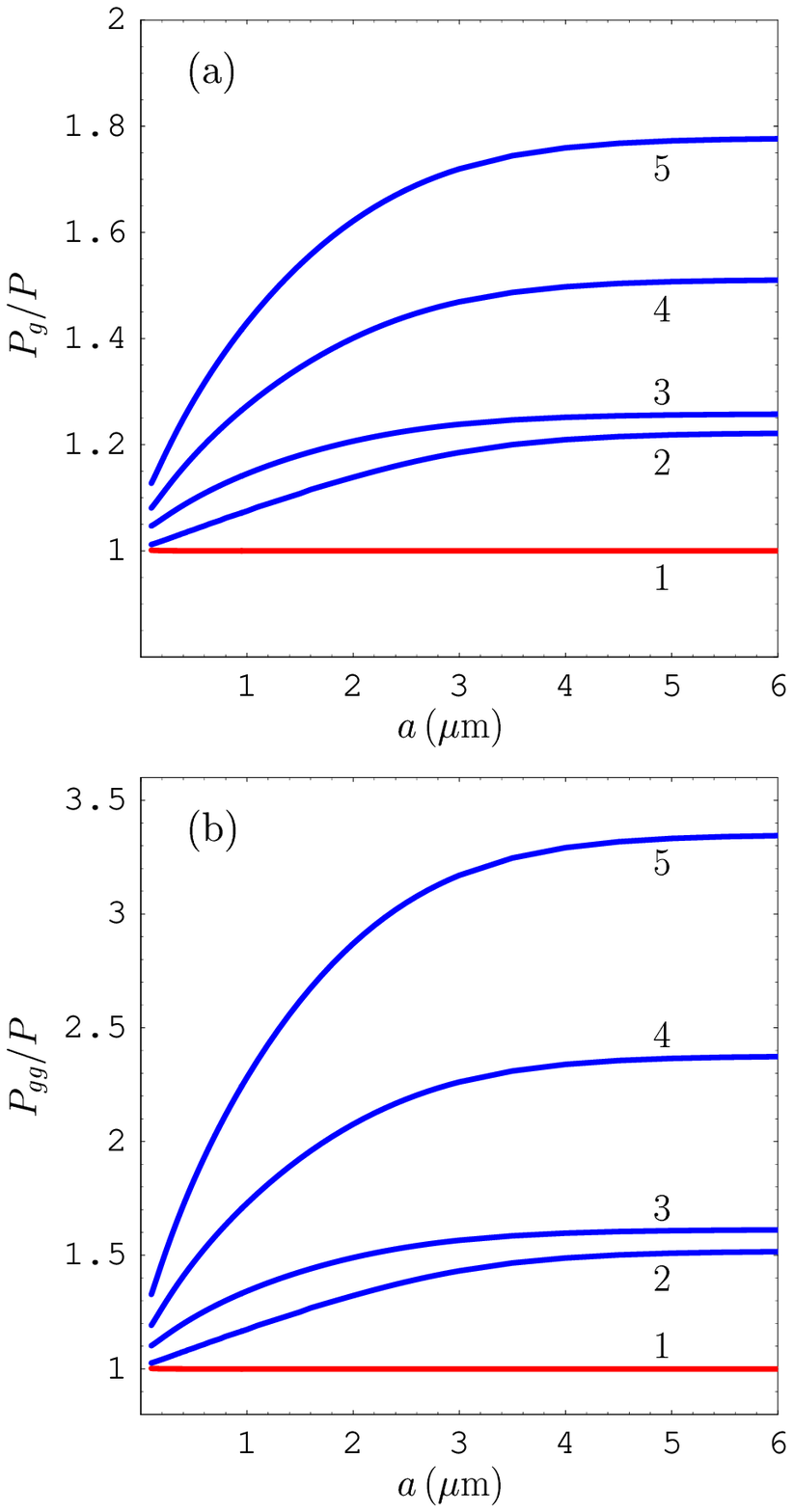}
}
\vspace*{-8.cm}
\caption{\label{fg1}(Color online)
Ratios of the Casimir pressures between two plates
(a) one of which is coated with graphene and (b) both
are coated with graphene to the Casimir pressure
between uncoated plates are shown as functions of
separation.
The lines numbered
1, 2, 3, 4, and 5  are for plate materials Au, Si,
sapphire, mica, and fused silica, respectively
}
\end{figure}
\begin{figure}[b]
\vspace*{-3cm}
\centerline{\hspace*{1cm}
\includegraphics{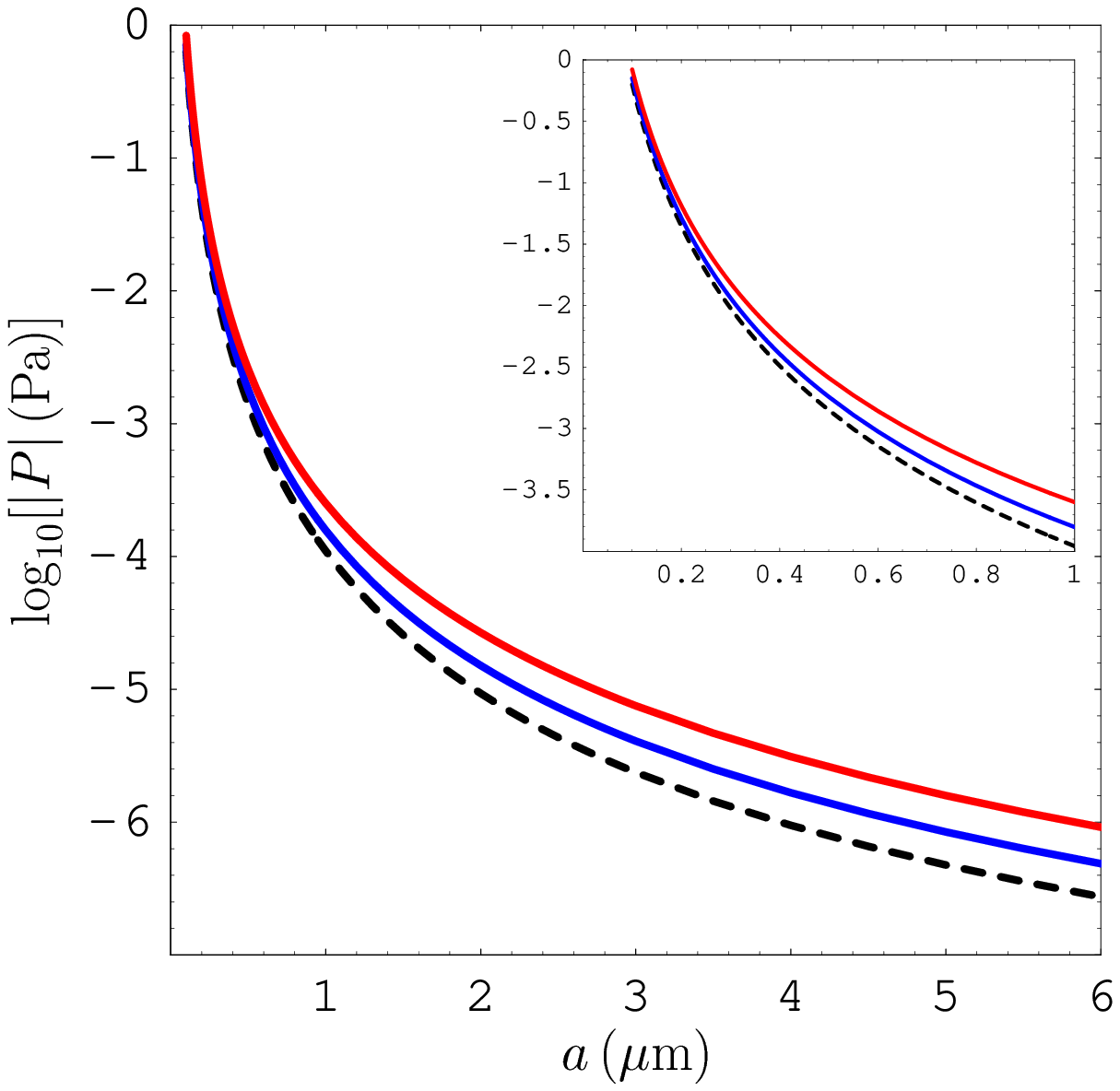}
}
\vspace*{-9.5cm}
\caption{\label{fg2}(Color online)
Magnitudes of the Casimir pressure between two fused silica
plates (dashed line), between one graphene-coated and one
uncoated (the lower solid line), and between two
graphene-coated fused silica plates (the upper solid line)
are plotted as functions of separation. The separation region
from 100\,nm to $1\,\mu$m is shown on an inset.
}
\end{figure}
\begin{figure}[b]
\vspace*{0.5cm}
\centerline{\hspace*{0.5 cm}
\includegraphics{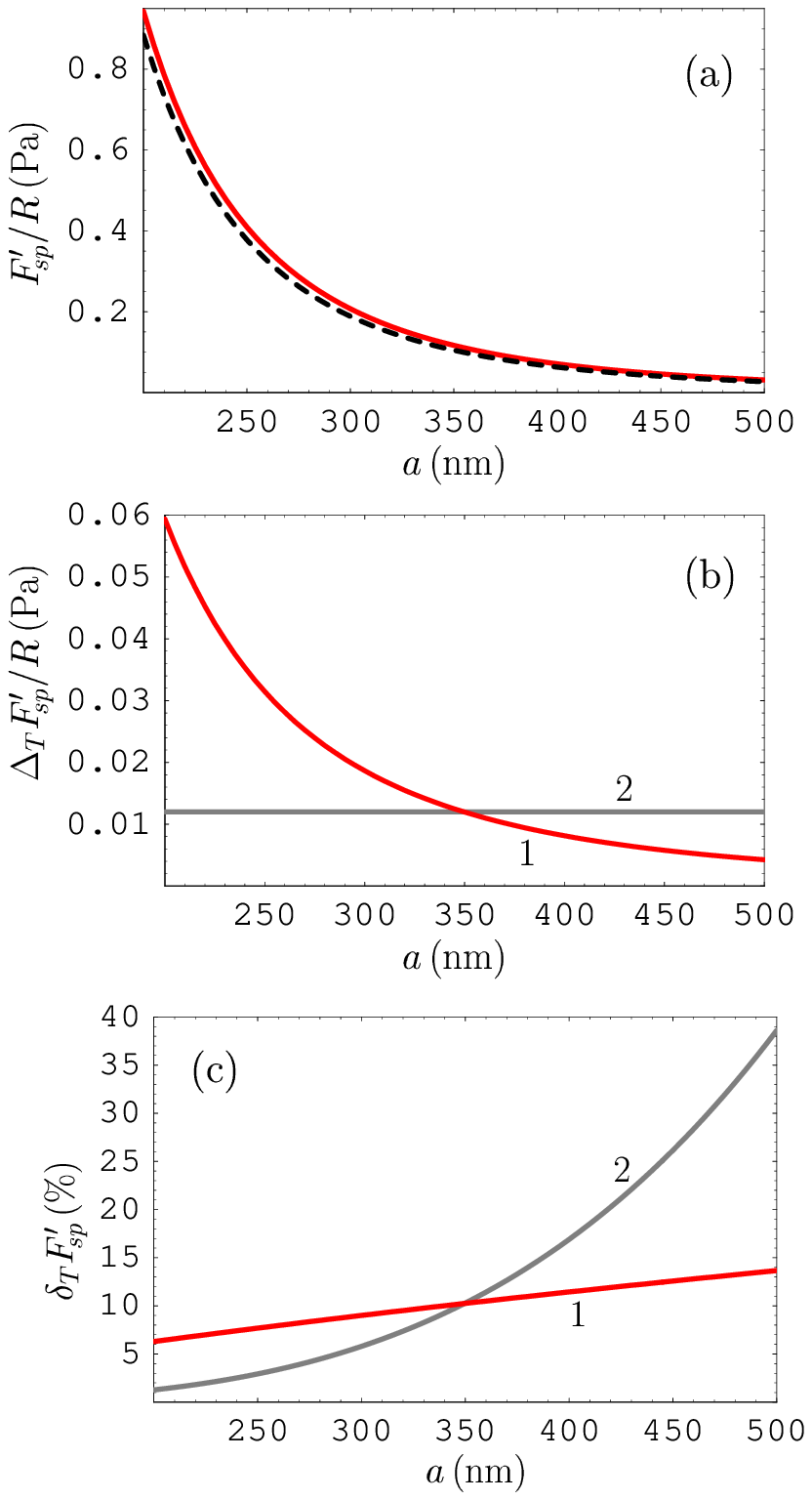}
}
\vspace*{-13cm}
\caption{\label{fg3}(Color online)
(a) The normalized gradient of the Casimir force
between an Au sphere of radius R and a graphene-coated
SiO${}_2$  plate at $T=300\,$K and $T=0\,$K are shown
by the solid and dashed lines, respectively.
The lines numbered 1 indicate (b) the absolute
and (c) the relative thermal correction to the
normalized force gradient. For comparison purposes,
the lines numbered 2 show  (b) the absolute
and (c) the relative total experimental errors
in the experiment \cite{17}.
}
\end{figure}
\begin{figure}[b]
\vspace*{-5cm}
\centerline{\hspace*{3cm}
\includegraphics{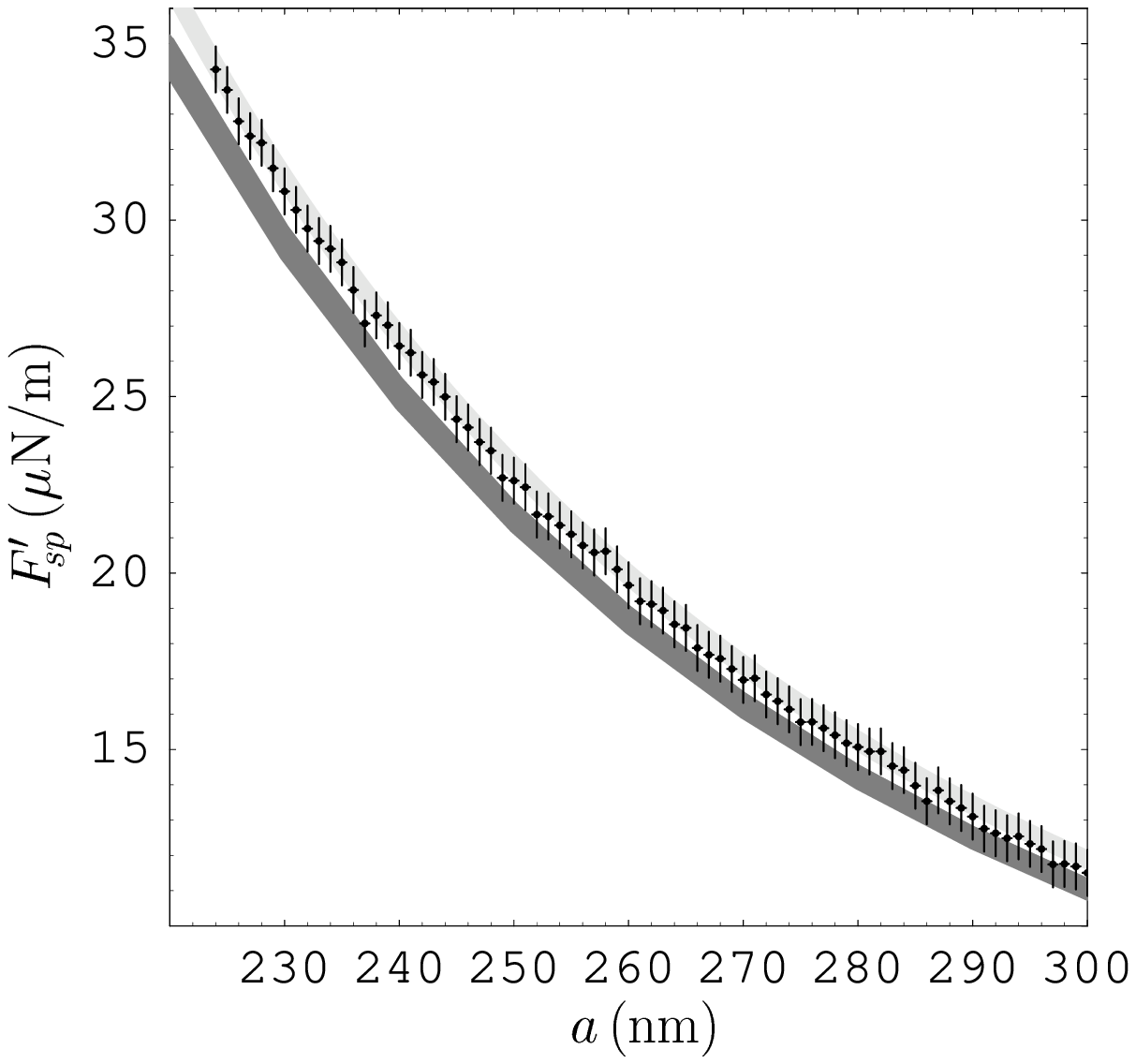}
}
\vspace*{-10cm}
\caption{\label{fg4}
The measured gradients of the Casimir force
between an Au-coated sphere and graphene deposited on a
SiO${}_2$ film covering a Si plate
are indicated as crosses at different separations.
The computational results at $T=300\,$K and $T=0\,$K are
shown by the light-gray and dark-gray bands, respectively.
See text for further discussion.
}
\end{figure}
\end{document}